\documentclass[english]{article}
\usepackage[T1]{fontenc}
\usepackage[latin9]{inputenc}
\usepackage{float}
\usepackage{textcomp}
\usepackage{amsmath}
\usepackage{amssymb}
\usepackage{graphicx}
\usepackage{babel}
\begin{document}

\title{Non-additive simple potentials for pre-programmed self-assembly}

\author{Daniel Salgado-Blanco and Carlos I. Mendoza%
\thanks{E-mail: cmendoza@iim.unam.mx%
}\\
 Instituto de Investigaciones en Materiales, Universidad Nacional\\
 Autónoma de México, Apdo. Postal 70-360, 04510 México, D.F., Mexico}
\maketitle
\begin{abstract}
A major goal in nanoscience and nanotechnology is the self-assembly
of any desired complex structure with a system of particles interacting
through simple potentials. To achieve this objective, intense experimental
and theoretical efforts are currently concentrated in the development
of the so called ``patchy'' particles. Here we follow a completely
different approach and introduce a very accessible model to produce
a large variety of pre-programmed two-dimensional (2D) complex structures.
Our model consists of a binary mixture of particles that interact
through isotropic in plane interactions that is able to self-assemble
into targeted lattices by the appropriate choice of a small number
of geometrical parameters and interaction strengths. We study the
system using Monte Carlo computer simulations and, despite its simplicity,
we are able to self assemble potentially useful structures such as
chains, stripes, Kagomé, twisted Kagomé, honeycomb, square, Archimedean
and quasicrystaline tilings. Our model is designed such that it may
be imediately implemented in experiments using existing techniques
to build particles with different shapes and interactions. Thus, it
represents a promising strategy for bottom-up nano-fabrication.
\end{abstract}

\section*{Introduction}

The quest for new materials with unusual physical properties and the
need to produce devices of technological interest at the nanoscale,
have boosted the design of new methods for the fabrication of complex
colloidal nanostructures. Processes such as micro- and nano-fabrication
are time consuming and prohibitively expensive, therefore they are
difficult to apply below a certain length scale \cite{choi}. As a
result, the search for building blocks on the mesoscopic scales that
self-organize into potentially useful structures by virtue of their
mutual interactions and shape is extremely important. One of the main
challenges is the ability to program the properties of the individual
components such that they organize into a desired structure \cite{grzybowski}.
In many cases this objective is pursued by trying to emulate the self-assembly
of living systems. Since most biomolecular objects interact through
directionally specific forces, a large amount of work has been done
to mimic the anisotropic nature of these interactions \cite{glotzer}-\cite{lee},
specifically, with the design and use of patchy \cite{zhang}-\cite{pawar}
and Janus \cite{jiang}-\cite{li-1} particles. This approach captures
much of the richness of nature's self-assembled structures and has
been successful in building some types of lattices \cite{chen}. However,
the production of particles with controlled patchiness in the laboratory
is still largely unavailable, although there has been impressive progress
in their synthesis \cite{lee}.

Different, mainly theoretical, procedures to tackle the problem are
the so-called inverse optimization techniques which consist in determining
the kind of isotropic interaction potential that would result in the
self-assembly of a desired structure \cite{torquato}-\cite{batten}.
Although this procedure has great potential, up to date, it results
in very complex interactions difficult to translate into a realistic
system. Multi-component colloidal systems interacting through simpler
isotropic potentials \cite{tang}-\cite{grunwald} are also an alternative
to build complex lattices. 

Nanometer-length-scale patterns in two dimensions are currently of
interest for its potential in many applications, such as optics, photonics,
sensing and others \cite{li-1}. Among the patterns that are of particular
interest, we can highlight the square lattice whose symmetry is appropriate
for using in nanocircuitry and therefore with prospective in the electronic
industry \cite{tang}, the Kagomé lattice for its applications in
the study of frustrated magnetism \cite{syozi}-\cite{mao}, or the
unusual mechanical properties like the auxetic response of the twisted
Kagomé lattices \cite{sun}, the honeycomb lattice for its electronic
properties motivated by its three-dimensional analog, the diamond
lattice \cite{torquato}, and the quasicrystals for photonics applications
\cite{roichman} among others.

\section*{Model}

\begin{figure}[H]

\begin{centering}
\includegraphics[scale=0.5]{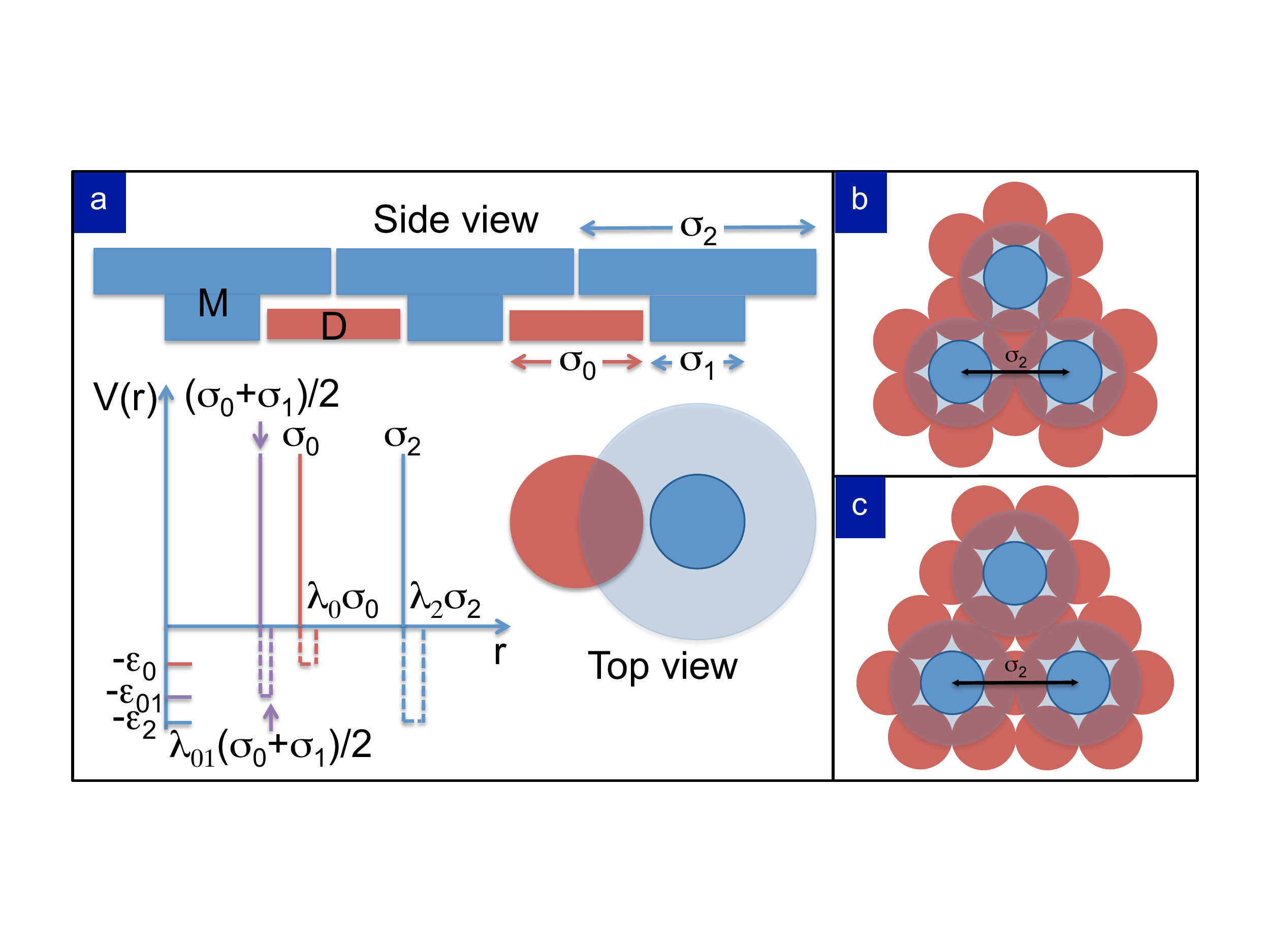} 
\par\end{centering}

\caption{\textbf{\small Description of the model.}{\small{} (a) Binary mixture
of mushroom-shaped particles $M$ (blue particles) and discs $D$
(red particles). The interaction potential between the hard discs
is depicted by the red line, the interaction between mushroom-shaped
particles is depicted by the blue line, and finally, the interaction
between a disk and a mushroom-shaped particle is depicted by the purple
line. The narrow attractive square well potential at the surface of
the particles is indicated by dashed lines. (b) and (c) Schematic
representation of two different lattices obtained for the same value
$\sigma_{1}/\sigma_{0}=1$. In both cases $M$-type particles lie
in a triangular lattice, however, in panel (b) the discs form a honeycomb
lattice for $\sigma_{2}/\sigma_{0}=\sqrt{3}$ while in panel (c) they
form a Kagomé lattice for $\sigma_{2}/\sigma_{0}=2$.}}

\label{figpotencial}
\end{figure}

Our system consists of a non-additive binary mixture of particles
as depicted in Fig. \ref{figpotencial}a. In a two-component mixture
normally the distance of closest approach between hard particles of
different species is a simple mean of the diameters of the particles
of each species. The non-additive hard particle mixture generalizes
this so that this distance can be smaller or larger than the arithmetic
mean of the like-species diameters \cite{hopkins}, \cite{faller}.
A 2D version of our model can be achieved as follows: one species
consists of two coupled layers of attractive hard discs as shown by
the mushroom-shaped particles ($M$) in Fig. \ref{figpotencial}a.
The second species consists of attractive hard discs ($D$) and both
species are able to move only in the plane perpendicular to their
symmetry axis. The interaction between particles is represented by
an axially symmetric pair potential $V(r)$ composed of an impenetrable
core surrounded by an adjacent square well. Our model is designed
to produce two-dimensional self-assembled structures in which $M$-type
particles are surrounded by discs such that each type of particles
arrange in mutually intercalated lattices. This methodology is particularly
useful for the self-assembly of open lattices. Since the lattices
of the two species are mutually intercalated, the open space of a
given lattice can be occupied by a particle of the second species,
thus providing stability to the structure during the formation process.
Discs have a core of diameter $\sigma_{0}$ and a thin square-well
potential with range $\lambda_{0}\sigma_{0}$. The interaction between
mushroom-shaped particles is represented by a core of diameter $\sigma_{2}$
and a thin square-well potential with range $\lambda_{2}\sigma_{2}$.
Finally, the interaction between a mushroom-shaped particle and a
disc consists of a hard core with diameter $\sigma_{01}=\left(\sigma_{0}+\sigma_{1}\right)/2$
and a thin adjacent square-well potential with range $\lambda_{01}\sigma_{01}$.
The non-additive nature of the model means that $\sigma_{01}=\left(\sigma_{0}+\sigma_{1}\right)/2=\left(1+\Delta\right)\left(\sigma_{0}+\sigma_{2}\right)/2$,
with $\Delta=\left(\sigma_{1}-\sigma_{2}\right)/\left(\sigma_{0}+\sigma_{2}\right)$.
The value of $\Delta$ in our model is always negative $\left(-1<\Delta\leq0\right)$
which means that the distance of closest approach between a disc and
a $M$-type particle is smaller than the mean of the diameters of
the particles of each species. The interaction potentials are also
depicted in Fig. \ref{figpotencial}a, where $\varepsilon_{0}$, $\varepsilon_{2}$,
and $\varepsilon_{01}$ are the depth of the potential wells. 

We study our system through Monte Carlo (MC) simulations at constant
number of particles $N$, volume $V$, and temperature $T$ ($NVT$
simulations). Our objective is to assemble different kinds of pre-programmed
structures in 2D, specifically, lattices with different symmetries
that are relevant for their scientific or technological interest.
The simplest lattice to assemble in 2D is the regular triangular lattice.
More difficult to assemble are open structures since they do not maximize
the translational entropy of the particles \cite{mao}. In our model
$M$-type particles are used as tool to produce open lattices made
of discs and viceversa. Among the many possible choices for the geometrical
parameters, one interesting possibility is to consider that each $M$-type
particle is surrounded by $n$ discs ($n\geq3$) closely packed around
the central $M$-type particle due to the attractive interaction $V_{DM}(r)$.
The value of $\sigma_{1}$ needed to allocate the discs is given by

\begin{equation}
\frac{\sigma_{1}}{\sigma_{0}}=\frac{\sqrt{2\left[1+\cos(\frac{2\pi}{n})\right]}}{\sin(\frac{2\pi}{n})}-1=\csc\left(\frac{\pi}{n}\right)-1.\label{sigma1}
\end{equation}
A given $M$-type particle may or may not share its surrounding discs
with other $M$-type particles. The way the discs are shared will
be determined by the value taken by $\sigma_{2}$ to finally produce
the desired lattice. For instance, in the tiling depicted in Fig.
\ref{figpotencial}b where $n=6$, the value $\sigma_{2}$ is chosen
so that each $M$-type particle shares two discs with each of its
neighboring $M$-type particles. On the other hand, in Fig. \ref{figpotencial}c,
even if each $M$-type particle is again surrounded by six discs,
the value chosen for $\sigma_{2}$ is such that each $M$-type particle
shares only one disc with each of its neighboring $M$-type particles.
Lattices similar to the one shown in Fig. \ref{figpotencial}b, with
each $M$-type particle surrounded by $n$ discs sharing two of them
with a neighboring $M$-type particle, can be constructed by choosing

\begin{equation}
\frac{\sigma_{2}}{\sigma_{0}}=\sqrt{\frac{\sigma_{1}}{\sigma_{0}}\left(2+\frac{\sigma_{1}}{\sigma_{0}}\right)}=\cot\left(\frac{\pi}{n}\right).\label{sigma2a}
\end{equation}
Steric interactions between discs restrict the use of Eq. (\ref{sigma2a})
to $n\leq12$.

On the other hand, for lattices similar to the one shown in Fig. \ref{figpotencial}c,
in which a $M$-type particle shares only one disc with a neighboring
$M$-type particle

\begin{equation}
\frac{\sigma_{2}}{\sigma_{0}}=1+\frac{\sigma_{1}}{\sigma_{0}}=\csc\left(\frac{\pi}{n}\right).\label{sigma2b}
\end{equation}
Steric interactions between discs restrict the use of Eq. (\ref{sigma2b})
to $n\leq6$.

Sometimes it is energetically more favorable for the system to phase
separate. To suppres this behavior, suitable values for the potential
wells should be chosen such that the discs prefer to stick around
a $M$-type particle.

Thus, the model can form a large variety of desired structures by
simply tailoring the geometrical parameters $\sigma_{1}/\sigma_{0}$,
and $\sigma_{2}/\sigma_{0}$, and the strength of the potential wells
$\varepsilon_{0}$, $\varepsilon_{2}$, and $\varepsilon_{01}$. The
width of the potential wells $\lambda_{0}$, $\lambda_{2}$, and $\lambda_{01}$
do not significantly alter obtained lattices and are only used for
fine tuning the resulting structure. The stoichiometry of the system
is determined by the lattice we desire to assemble.

\section*{Results and discussion}

\begin{figure}[H]
\centering{}{\footnotesize \includegraphics[bb=120bp 0bp 610bp 540bp,scale=0.7]{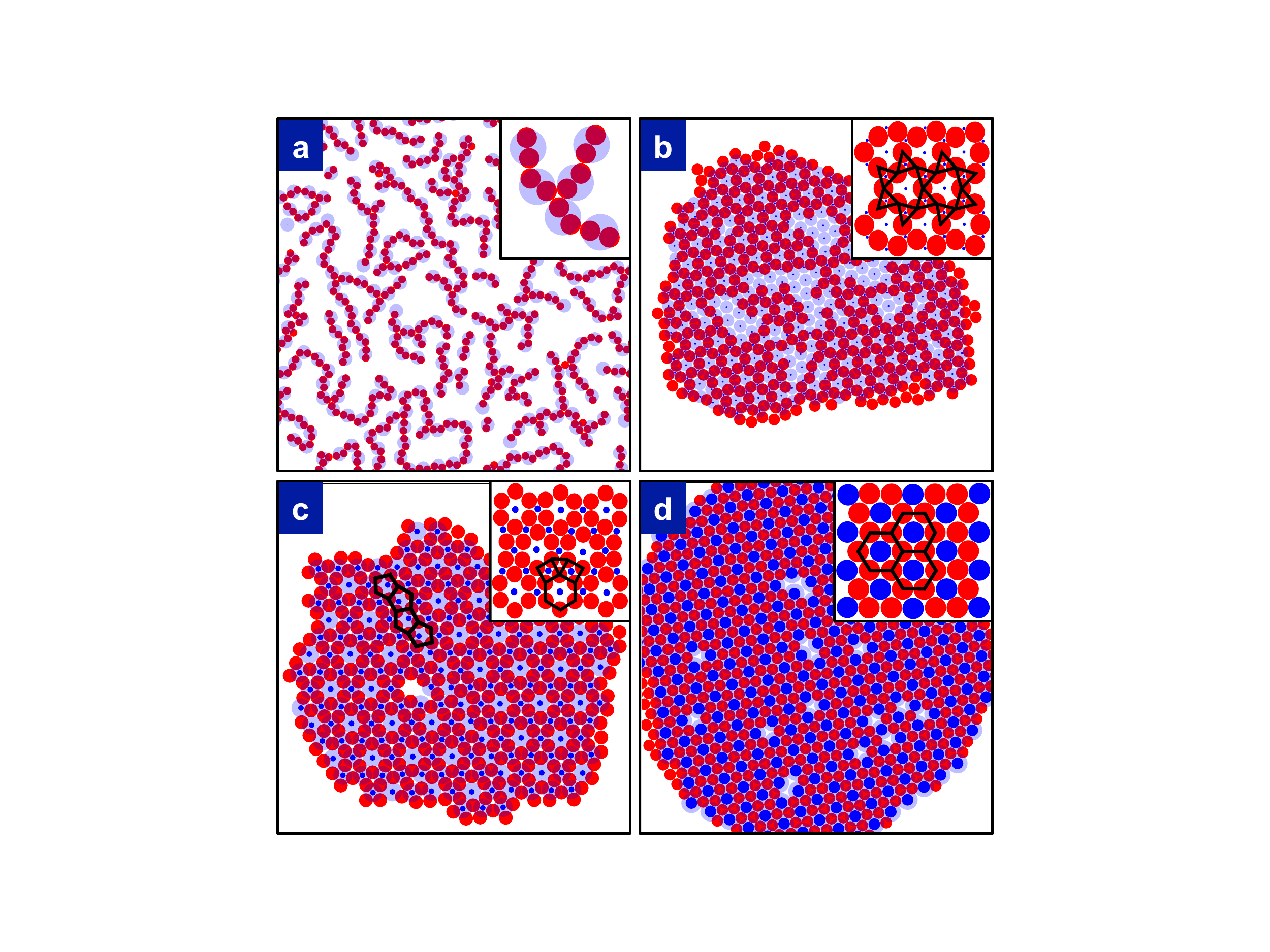}}\caption{\textbf{\small Chains, twisted Kagomé, Archimedean, and honeycomb
tilings.}{\small{} (a) Polymer like structures obtained with $\sigma_{1}/\sigma_{0}=0.02$
and $\sigma_{2}/\sigma_{0}=1.8$. A few branching points are present
as shown in the inset. (b) Twisted Kagomé lattice obtained with $n=3$,
$\sigma_{1}/\sigma_{0}\simeq0.1547$, as given by Eq. (\ref{sigma1})
and $\sigma_{2}/\sigma_{0}=1.1$. The inset shows two plaquetes of
the lattice. (c) Semi-regular rhombitrihexagonal tiling of discs (red)
obtained with $n=4$, $\sigma_{1}/\sigma_{0}=\sqrt{2}-1$, as given
by Eq. (\ref{sigma1}), and $\sigma_{2}/\sigma_{0}\simeq1.37$. The
lattice is intercalated with a triangular lattice of $M$-type particles
(blue). A dislocation line in the lattice of discs formed by pentagons
is highlighted with black lines. The inset shows the $\left(3.4.6.4\right)$
vertex that decorates the Archimedean lattice of discs. (d) Honeycomb
lattice of discs (red) obtained with $n=6$, $\sigma_{1}/\sigma_{0}=1$,
as given by Eq. (\ref{sigma1}), and $\sigma_{2}/\sigma_{0}\simeq1.73$,
as given by Eq. (\ref{sigma2a}). The depth of the potential wells
$\left(\varepsilon_{0},\varepsilon_{2},\varepsilon_{01}\right)$ for
each structure are $\left(0.5,0,1.5\right)$, $\left(0.7,1,1\right)$,
$\left(0.5,1.5,1\right)$ and $\left(1,1,1\right)$, respectively.}}
\label{fign234}
\end{figure}

In what follows we explore the parameter space set by the parameters
$\sigma_{1}/\sigma_{0}$, $\sigma_{2}/\sigma_{0}$, $\varepsilon_{0}$,
$\varepsilon_{2}$, and $\varepsilon_{01}$, and construct a number
of different target structures. First, we consider the case in which
each $M$-type particle is in contact with only two discs in order
to form chains. We can achieve this by setting a small value of $\sigma_{1}/\sigma_{0}$.
One example is displayed in Fig. \ref{fign234} panel (a). It shows
the result for $\sigma_{1}/\sigma_{0}=0.02$, $\sigma_{2}/\sigma_{0}=1.8$,
and $\varepsilon_{2}=0$. The small value for $\sigma_{1}/\sigma_{0}$
is chosen so that $M$-type particles act as stickers between two
discs. On the other hand, the value of $\sigma_{2}/\sigma_{0}$ is
chosen so that only a small fraction of the discs protrudes from the
cap of the $M$-type particles therefore forming effectively anisotropic
particles with two interacting patches. The resulting patchy particles
join to form flexible chains with a few branching points. Furthermore,
by varying the values of $\sigma_{1}/\sigma_{0}$ and $\sigma_{2}/\sigma_{0}$,
the persistence length of the chains can be controlled to certain
extent. Other chain structures and stripes are shown in Figs. \ref{suppfign23468}a
and b.

Now we turn to the cases given by Eq. (\ref{sigma1}), progressively
increasing the value of $n$ and using different choices for $\sigma_{2}/\sigma_{0}$
and the depth of the potential wells, $\varepsilon_{0}$, $\varepsilon_{2}$,
and $\varepsilon_{01}$. For $n=3$, using Eqs. (\ref{sigma1}) and
(\ref{sigma2a}), a triangular lattice made of discs is intercalated
with a honeycomb lattice made of $M$-type particles, as shown in
Fig. \ref{suppfign23468}c. On the other hand, if Eq. (\ref{sigma2b})
is used, then a Kagomé lattice of discs is intercalated with a triangular
lattice of $M$-type particles, as shown in Fig. \ref{suppfign23468}d.
Kagomé lattices have been self-assembled using trijanus particles
\cite{chen} and its stability in this case is favored by entropy
\cite{mao}. However, in our model the relevant quantity is energy
since the system is trying to minimize their interactions by maximizing
the number of favorable contacts between particles.

Another interesting target structure with a great deal of technological
potential is the twisted Kagomé lattice \cite{torquato} since it
is an arrangement that presents negative Poisson's ratio (auxetic
behavior) \cite{sun}. An auxetic material, when stretched in a particular
direction, expands in an orthogonal direction. In the present model,
twisted Kagomé lattices are obtained for intermediate values of $\sigma/\sigma_{0}$
as shown in Fig. \ref{fign234}b, where the value $\sigma_{2}/\sigma_{0}=1.1$
is used. It has been shown that twisted Kagomé lattices can be obtained
as a minimum energy configuration of patchy particles with five-patch
particles, decorated with two $A$ and three $B$ patches, in which
like patches attract each other, while unlike patches repel each other
\cite{doppelbauer}. In contrast, in our model, the twisted Kagomé
lattices are self-assembled using only isotropic (in the plane containing
the particles) interactions.

An example of lattice obtained with $n=4$ is shown in Fig. \ref{fign234}
panel (c). It shows a triangular lattice of $M$-type particles intercalated
with a very open structure of discs. Lines connecting neighboring
discs show that each vertex of the lattice is surrounded by a triangle,
two squares and an hexagon {[}inset of Fig. \ref{fign234} panel (c){]}.
In general, the vertex of a tiling made of regular polygons can be
described as $\left(n_{1}.n_{2}.n_{3}...\right)$ corresponding to
the numbers of sides of the polygons listed in order. Using that notation,
our lattice can be written as $\left(3.4.6.4\right)$. This lattice
is known as semi-regular, rhombitrihexagonal tiling and is an example
of Archimedean tiling {[}inset of Fig. \ref{fign234} panel (c){]}.
Archimedean tilings are defined as regular patterns of polygonal tessellation
of a plane by regular polygons where only one type of vertex is permitted
in each tiling. Such Archimedean tilings have recently been self assembled
using enthalpically and entropically patchy polygons \cite{millan}.
Notice the interesting dislocation consisting of a chain of pentagons
as it is highlighted by the black lines in Fig. \ref{fign234}c. Lattices
with square symmetry obtained with $n=4$ are shown in Figs. \ref{suppfign23468}e
and f.

\begin{figure}[H]
\centering{}\includegraphics[bb=120bp 0bp 720bp 540bp,scale=0.7]{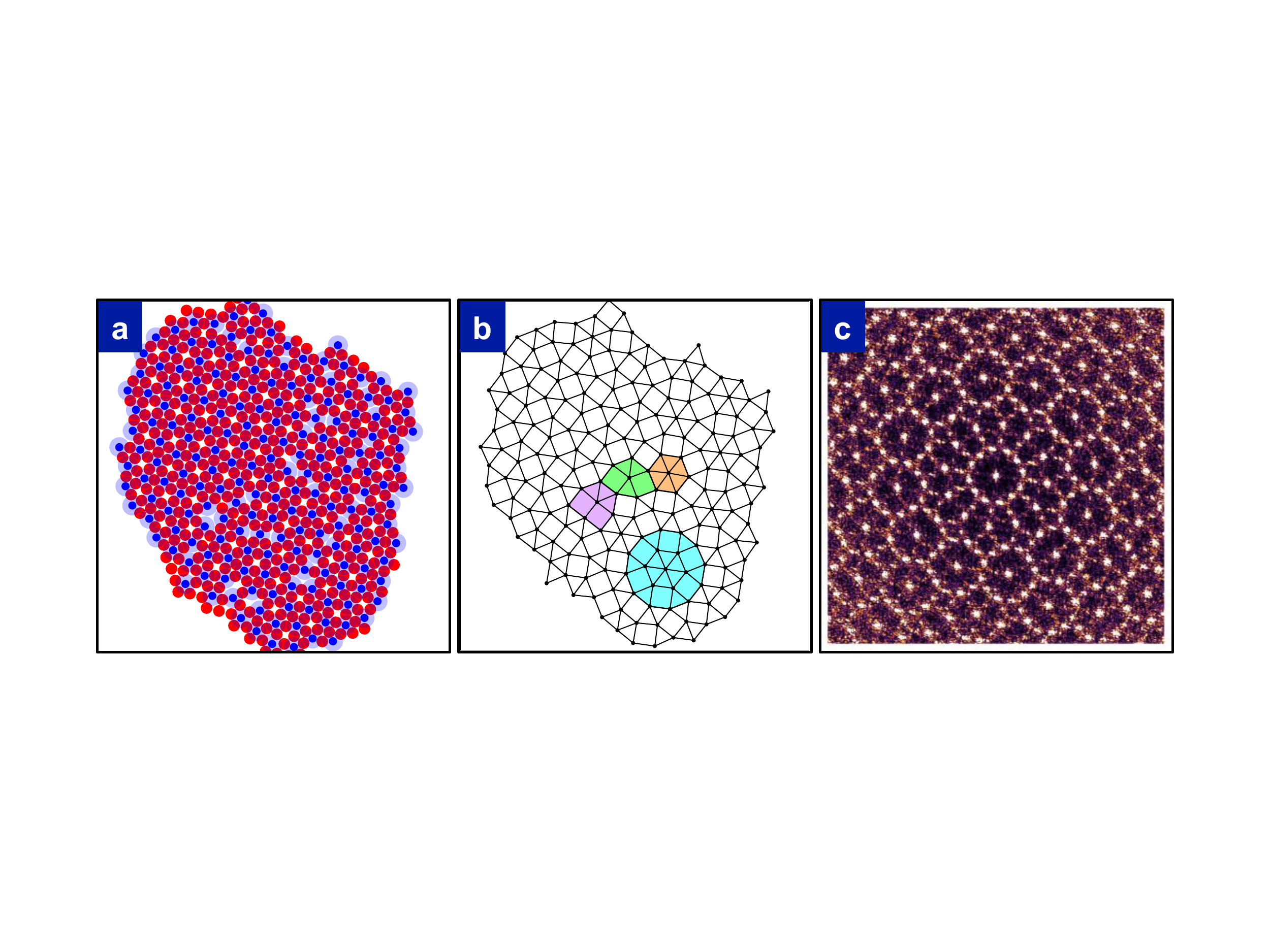}\caption{\textbf{\small Dodecagonal quasicrystal.}{\small{} (a) Structure obtained
with $n=5$, $\sigma_{1}/\sigma_{0}\simeq0.7013$, as given by Eq.
(\ref{sigma1}), and $\sigma_{2}/\sigma_{0}\simeq1.72$. In this case
a dodecagonal quasicrystal of $M$-type particles (blue) is intercalated
with a lattice of pentagons made of discs (red).$ $ The structure
has a twelvefold symmetry. (b) Square-triangular pattern corresponding
to the $M$-type particles. Three neighbor classification of $\sigma$
(green), $H$ (purple), and $Z$ (orange) environments are shown.
A dodecagonal motif tipically found in quasicrystals is highlighted
in cyan. (c) Diffraction pattern of the lattice formed by the $M$-type
particles showing dodecagonal symmetry. The depth of the potential
wells is $\left(\varepsilon_{0},\varepsilon_{2},\varepsilon_{01}\right)=\left(1,1,1\right)$.}}
\label{fign5}
\end{figure}

Clearly, the case with $n=5$ is particularly interesting since in
this case the local symmetry is incompatible with crystalline order.
This suggests the possibility to construct aperiodic structures with
long-range order, that is, quasicrystals (or their approximants).
Quasicrystaline heterostructures fabricated from dielectric materials
with micrometer-scale features exhibit interesting and useful optical
properties including large photonics bandgaps in two-dimensional systems
\cite{roichman}. Thus, they are an interesting case to self-assemble.
As expected, it is possible to choose the geometrical parameters such
that the resulting structure present rotational symmetry consistent
with a twelvefold-symmetric quasicrystal as shown in Fig. \ref{fign5}a.
Lines connecting neighboring $M$-type particles of the whole lattice
show a square-triangular tiling (see Fig. \ref{fign5}b) whose vertices
can be of three different types: $\left(3^{2}.4.3.4\right)$ is highlighted
with green color, $\left(3^{3}.4^{2}\right)$ is marked in purple,
and $\left(3^{6}\right)$ in orange. A dodecagonal structural motif
usually present in quasicrystals is shown in cyan color. It is known,
that patterns of squares and triangles tend to form twelvefold-symmetric
quasicrystals \cite{oxborrow},\cite{widom}. A confirmation of this
fact is the diffraction pattern of the $M$-type particles lattice
which is consistent with a dodecagonal quasicrystal, as shown in Fig.
\ref{fign5}c. Alternative procedures to self-assemble quasicrystals
and their approximants have been proposed, they are based on particle
functionalization with mobile surface entities and shape polydispersity
\cite{iacovella} or with the use of five and seven patched particles
\cite{vanderlinden}. In contrast, our method uses only isotropic
interactions.

The case with $n=6$ provides an alternative procedure to construct
the honeycomb and Kagomé lattices. The first case, obtained using
Eqs. (\ref{sigma1}) and (\ref{sigma2a}) is shown in Fig. \ref{fign234}d.
Other configurations are shown in Figs. \ref{suppfign23468} to \ref{suppfign5}.

\begin{figure}[H]
\centering{}\includegraphics[bb=120bp 0bp 720bp 540bp,scale=0.7]{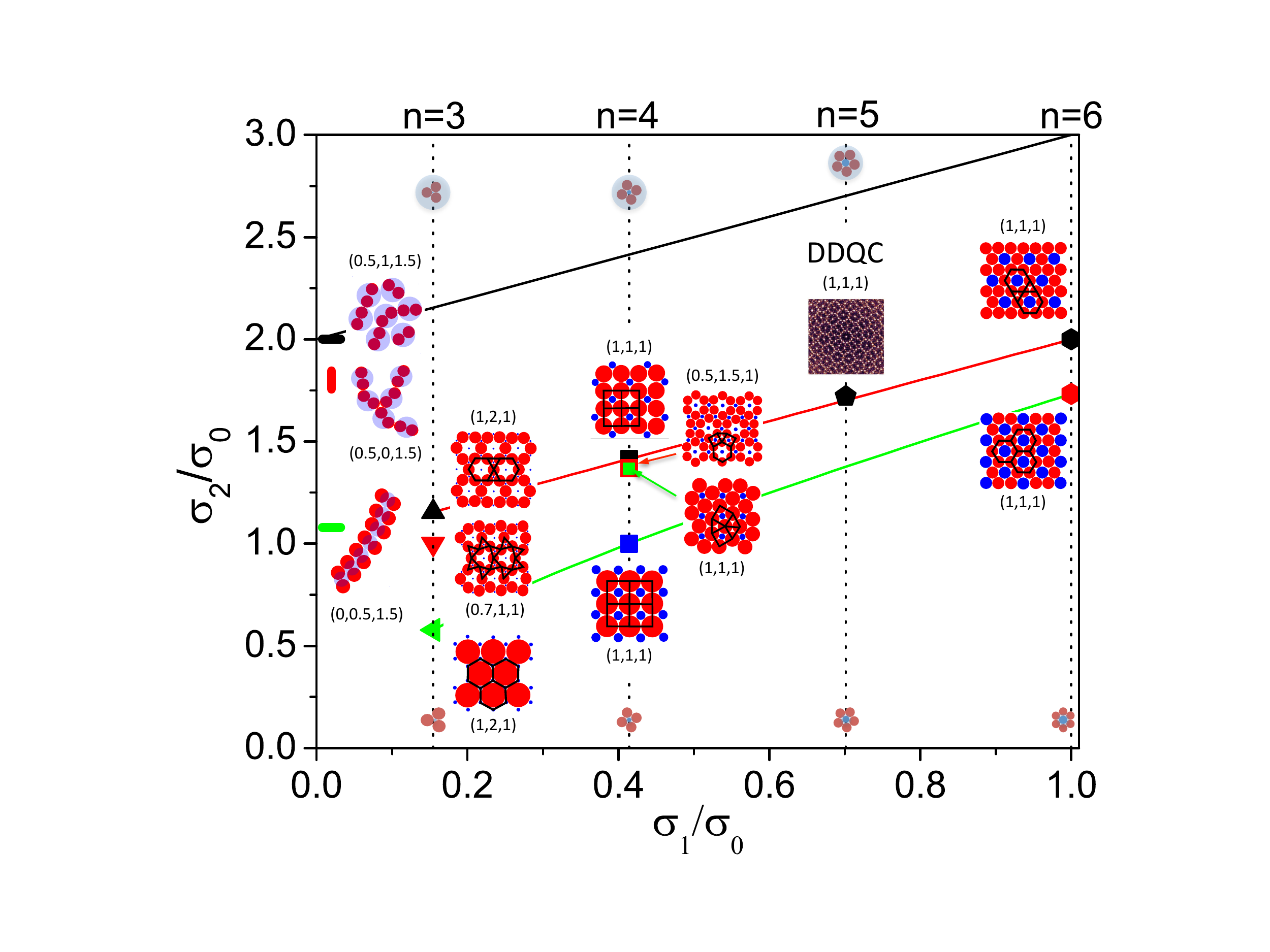}\caption{\textbf{\small Phase diagram.}{\small{} Summary of the self-assembled
structures. The green and red lines are the values of $\sigma_{2}/\sigma_{0}$
given by Eqs. (\ref{sigma2a}) and (\ref{sigma2b}), respectively.
The black straight line is $\sigma_{2}/\sigma_{0}=2+\sigma_{1}/\sigma_{0}$.
Above this value, the system consists of a fluid (if $\varepsilon_{2}=0$
and for low concentrations) or a crystal (if $\varepsilon_{2}\neq0$
or for large concentrations) of meta-particles. Symbols correspond
to the structures built in this study and the triplets $\left(\varepsilon_{0},\varepsilon_{2},\varepsilon_{01}\right)$
above or below each inset correspond to the energies used to obtain
the corresponding lattice. The dodecagonal quasicrystal (DDQC) is
represented by its diffraction pattern.}}
\label{figphasediag}
\end{figure}

Our results are summarized in the zero temperature phase diagram shown
in Fig. \ref{figphasediag}. The green and red lines represent Eqs.
(\ref{sigma2a}) and (\ref{sigma2b}), respectively. The energies
used to obtain any given structure are indicated by the triplets $\left(\varepsilon_{0},\varepsilon_{2},\varepsilon_{01}\right)$.
Clearly, for the same set of {\small $\sigma_{1}/\sigma_{0}$} and
{\small $\sigma_{2}/\sigma_{0}$}, other structures could be obtained
by using different choices for the energies. Above the straight-line
$\sigma_{2}/\sigma_{0}=2+\sigma_{1}/\sigma_{0}$ {\small and for the
right }stoichiometry,{\small{} the system consists of a fluid (if $\varepsilon_{2}=0$
and for low concentrations) or a crystal (if $\varepsilon_{2}\neq0$
or for large concentrations) of meta-particles composed of a $M$-type
particle surrounded by $n$ discs. The large value of $\sigma_{2}$
prevents the interaction of discs belonging to different meta-particles,
therefore the meta-particles interact as isotropic discs of diameter
$\sigma_{2}$. On the other hand, for values of $\sigma_{2}/\sigma_{0}$
below the given by Eq. }(\ref{sigma2a}), {\small the meta-particle
interactions have $n$-gonal symmetry, as represented by the drawings
in Fig. \ref{figphasediag}.}{\small \par}

\section*{Conclusions}

In conclusion, we have presented a very simple model that is able
to generate a large variety of pre-programmed structures. We emphasize
the simplicity of the interactions which are isotropic in the plane
containing the particles, and the relative ease with which we get
complex structures by controlling a small number of geometric and
energetic parameters. Furthermore, our two species model can be straightforwardly
generalized to three or more species to construct more complex lattices,
including, for example, self-similar structures. Let us stress that
the simplicity of the model, the precise control in current nanotechnology
to produce particles with different shapes and the large variety of
methods to produce short range attractions, including the use of depletion
\cite{lekkerkerker} or DNA-mediated interactions \cite{kim}, make
it very realistic the feasibility to put into practice the present
model. Finally, we suggest that if the experiments are made using
chemically or temperature sensitive particles that can change size,
then the system could potentially switch smoothly between different
lattices, something that would be difficult to achieve with other
systems.

\section*{Acknowledgements}

We are grateful to Zorana Zeravcic for useful comments. This work
was supported in part by grant DGAPA IN-110613. DSB acknowledges financial
support from CONACyT through scholarship Num. 207347.

\section*{Appendix A: Methods}

Standard Monte Carlo (MC) simulations based on the canonical ensemble
(NVT simulations) in a square box of side $L$ with periodic boundary
conditions have been carried out using the Metropolis algorithm. We
have used $\sigma_{0}$ and $\varepsilon_{0}$ as length and energy
units, respectively, the reduced temperature $T^{*}=k_{B}T/\varepsilon_{0}$,
where $k_{B}$ is Boltzmann\textasciiacute{}s constant; the reduced
number density $\rho^{*}=\left(N_{M}\sigma_{1}^{2}+N_{D}\sigma_{0}^{2}\right)/L^{2}$,
where $N_{i}$ stands for the number of particles of species $i$.

Simulations were performed with $N\approx1000$ particles, and control
runs with $N=5000$ particles to exclude finite size effects were
also done. In all cases, the system is first disordered at high temperature
and then brought from $T^{*}=3.0$ to the final temperature $T^{*}=0.01$
through an accurate annealing procedure with steps of $0.01$. An
equilibration cycle consisted, for each temperature, of at least $1\times10^{8}$
MC steps, each one representing one trial displacement of each particle,
on average. At every simulation step a particle is picked at random
and given a uniform random trial displacement within a radius of $0.1\sigma_{0}$.
The range of the potential wells were $\lambda_{i}=1.05\sigma_{i}$,
with $i=0,2,01$.

\section*{Appendix B: Additional discussion}

Mathematically, the interaction potentials can be expressed by the
following set of equations

\[
V_{DD}(r)=\left\{ \begin{array}{l}
\infty,\text{ \ \ \ \ \ if }r\leq\sigma_{0}\\
-\varepsilon_{0},\text{ \ \ \ \ if }\sigma_{0}<r\leq\lambda_{0}\sigma_{0}\\
0,\text{ \ \ \ \ \ \ if }r>\lambda_{0}\sigma_{0}
\end{array}\right.,
\]

\[
V_{MM}(r)=\left\{ \begin{array}{l}
\infty,\text{ \ \ \ \ \ if }r\leq\sigma_{2}\\
-\varepsilon_{2},\text{ \ \ \ \ if }\sigma_{2}<r\leq\lambda_{2}\sigma_{2}\\
0,\text{ \ \ \ \ \ \ if }r>\lambda_{2}\sigma_{2}
\end{array}\right.,
\]

\[
V_{DM}(r)=\left\{ \begin{array}{l}
\infty,\text{ \ \ \ \ \ if }r\leq\frac{(\sigma_{0}+\sigma_{1})}{2}\\
-\varepsilon_{01},\text{ \ \ \ \ if }\frac{(\sigma_{0}+\sigma_{1})}{2}<r\leq\lambda_{01}\frac{(\sigma_{0}+\sigma_{1})}{2}\\
0,\text{ \ \ \ \ \ \ if }r>\lambda_{01}\frac{(\sigma_{0}+\sigma_{1})}{2}
\end{array}\right.,
\]
where $V_{ij}$ represents the interaction potential between a particle
$i=D,M$ and a particle $j=D,M$. The distance between the central
axes of the particles is $r$.

\begin{figure}[H]
\centering{}{\footnotesize \includegraphics[scale=0.5]{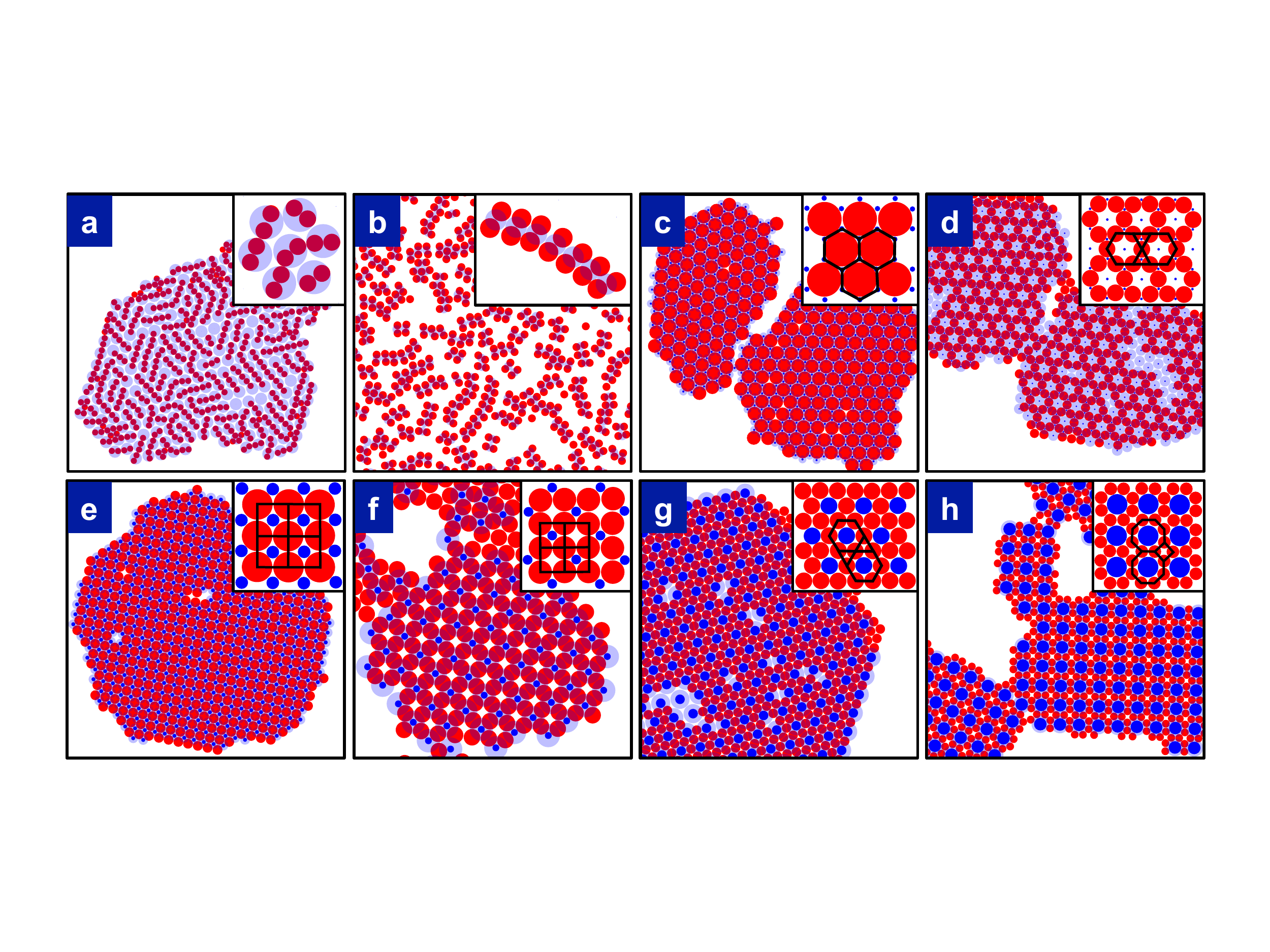}}\caption{\textbf{\small Miscellaneous structures.}{\small{} Structures obtained
with (a) $\sigma_{1}/\sigma_{0}=0.02$ and $\sigma_{2}/\sigma_{0}=2$.
(b) $\sigma_{1}/\sigma_{0}=0.02$ and $\sigma_{2}/\sigma_{0}=1.08$.
(c) $n=3$, $\sigma_{1}/\sigma_{0}$, and $\sigma_{2}/\sigma_{0}$
as given by Eqs. (1) and (2), respectively. A triangular lattice of
discs (red) is intercalated with a honeycomb lattice of $M$-type
particles (blue). The inset shows the $\left(6^{3}\right)$ motif
of the regular $M$-type particle tiling. (d) $n=3$, $\sigma_{1}/\sigma_{0}$,
and $\sigma_{2}/\sigma_{0}$ as given by Eqs. (1) and (3). A triangular
lattice of $M$-type particles (blue) is intercalated with a Kagomé
lattice of discs (red).$ $The inset shows the $\left(3.6.3.6\right)$
motif of the disc tiling. (e) $n=4$, $\sigma_{1}/\sigma_{0}$, and
$\sigma_{2}/\sigma_{0}$ as given by Eqs. (1) and (2). In this case
a square lattice of discs (red) is intercalated with a square lattice
of $M$-type particles (blue). (f) $n=4$, $\sigma_{1}/\sigma_{0}$,
and $\sigma_{2}/\sigma_{0}$ as given by Eqs. (1) and (3). In this
case a square lattice of discs (red) is intercalated with a square
lattice of $M$-type particles (blue) rotated 45 degrees with respect
to the first lattice. (g) $n=6$, $\sigma_{1}/\sigma_{0}$, and $\sigma_{2}/\sigma_{0}$
as given by Eqs. (1) and (3). A triangular lattice of $M$-type particles
(blue) is intercalated with a Kagomé lattice of discs (red). The inset
shows the $\left(3.6.3.6\right)$ motif of the Kagomé tiling. (h)
$n=8$, $\sigma_{1}/\sigma_{0}$, and $\sigma_{2}/\sigma_{0}$ as
given by Eqs. (1) and (2). In this case a square lattice of $M$-type
particles (blue) is intercalated with a truncated square tiling of
discs (red). The inset shows the $\left(4.8^{2}\right)$ motif. The
depths of the potential wells for each case are indicated in the phase
diagram, Fig. 4.}}
\label{suppfign23468}
\end{figure}

Fig. \ref{suppfign23468} Panel (a) shows the result of using $\sigma_{1}/\sigma_{0}=0.02$,
$\sigma_{2}/\sigma_{0}=2$ and $\left(\varepsilon_{0},\varepsilon_{2},\varepsilon_{01}\right)=\left(0.5,1,1.5\right)$.
In this case, $M$-type particles form a triangular lattice to maximize
their favorable contacts and the discs accommodate in domains of mostly
parallel stripes, some of them with a few bends. Panel (b) shows the
result when $\sigma_{1}/\sigma_{0}=0.02$, $\sigma_{2}/\sigma_{0}=1.08$,
and $\left(\varepsilon_{0},\varepsilon_{2},\varepsilon_{01}\right)=\left(0,1,1\right)$.
Panel (c) shows the case for $n=3$ using Eqs. (1) and (2), a triangular
lattice made of discs is intercalated with a honeycomb lattice made
of $M$-type particles. Drawing lines joining each particle of the
lattice with their nearest neighbors we observe that it can be characterized
by a plane tiling of regular hexagons (inset). Panel (d) shows another
case with $n=3$. Using Eqs. (1) and (3) a Kagomé lattice of discs
is intercalated with a triangular lattice of $M$-type particles.
In the Kagomé lattice each particle is in contact with four other
particles of the same species. If we tessellate the Kagomé lattice
by drawing lines between nearest neighbors, we observe that each vertex
can be written as $\left(3.6.3.6\right)$ and therefore is also known
as trihexagonal tiling (see inset). Notice that in this case not all
$M$-type particles are equivalent since some of them are in contact
with three discs while others, located at the pores of the Kagomé
lattices are not in contact with the discs but only with neighboring
$M$-type particles. Panel (e) shows a case with $n=4$, using Eqs.
(1) and (2). Two intercalated square lattices are formed. Panel (f)
shows the structure obtained when using Eq. (3). Two square lattices
are formed but their principal axis are rotated $45$ degrees with
respect to each other, and a square lattice of voids is also apparent.
The case with $n=6$ provides an alternative procedure to construct
the Kagomé lattice. When using Eq. (3) a triangular lattice of $M$-type
particles intercalated with a Kagomé lattice of discs is obtained
{[}panel (g){]}. We have not obtained regular lattices or other recognizable
structures formed with $n=7$. Finally, the structure formed with
$n=8$ and using Eqs. (1) and (2), is shown in panel (h). The truncated
square tiling with vertex $\left(4.8^{2}\right)$ is shown in the
inset.

\begin{figure}[H]
\begin{centering}
\includegraphics[scale=0.5]{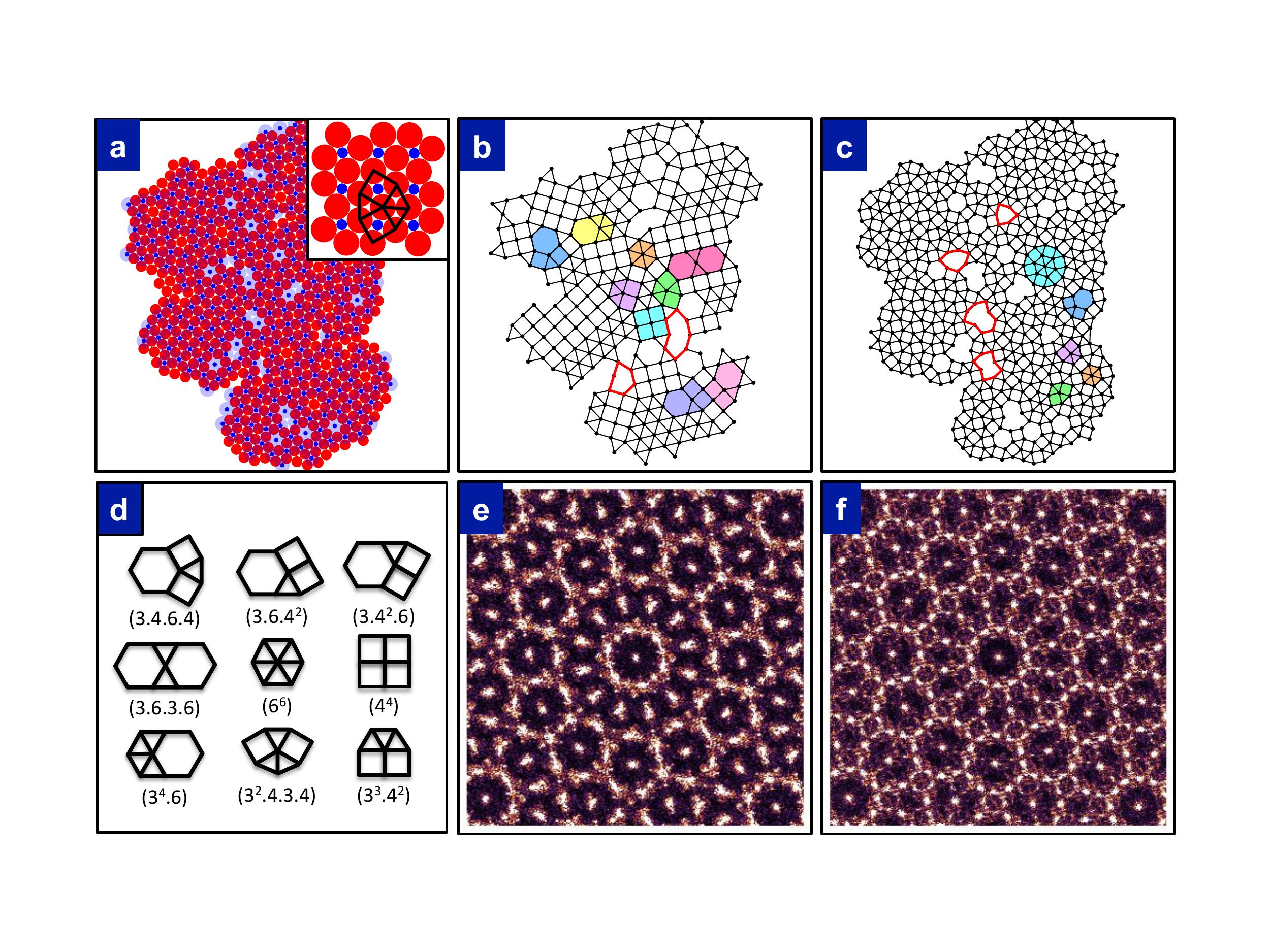}
\par\end{centering}

\centering{}\caption{\textbf{\small Polycrystalline snub square.}{\small{} (a) Polycrystalline
snub square lattice of discs (red) obtained with $n=4$, $\sigma_{1}/\sigma_{0}$,
as given by Eq. (1) and $\sigma_{2}/\sigma_{0}\simeq1.366$. For a
given domain (see inset), the lattice of discs is intercalated with
a square lattice of $M$-type particles (blue). The inset shows the
$\left(3^{2}.4.3.4\right)$ vertex that decorate the lattice of discs.
(b) Square-triangular pattern corresponding to the $M$-type particles.
The different vertices forming this pattern are highlighted with shadowed
tiles. The type of defects present in the structure are highlighted
with red lines. (c) Square-triangular pattern corresponding to the
discs. The different vertices forming this pattern are highlighted
with shadowed tiles. The type of defects present in the structure
are highlighted with red lines. A dodecagonal motif typically found
in quasicrystals is highlighted in cyan. (d) Summary of the types
of vertices found in the patterns shown in panels (b) and (c). (e)
and (f) diffraction patterns of the lattice of $M$-type particles
and of discs, respectively, showing twelve-fold symmetry. The depths
of the potential wells are $\left(1,1,1\right)$.}}
\label{suppfign4}
\end{figure}

Fig. \ref{suppfign4} shows a case obtained with $n=4$. Eq. (1) gives
$\sigma_{1}/\sigma_{0}\simeq0.4142$ and we have used $\sigma_{2}/\sigma_{0}\simeq1.366$
and $\left(\varepsilon_{0},\varepsilon_{2},\varepsilon_{01}\right)=\left(1,1,1\right)$.
In panel (a) we observe that $M$-type particles form a regular square
lattice while each disc is in contact with five other discs and form
a lattice of ``tilted'' squares. Drawing lines connecting neighboring
discs we observe that each vertex of the lattice can be written as
$\left(3^{2}.4.3.4\right)$, a lattice also known as snub square tiling
(see inset). In panels (b) and (c) we show the structural motifs formed
by joining with lines neighboring particles that compose the $M$-type
particle and disc lattices, respectively. Note the defects present
in the lattices, basically, vacancies with different geometrical shapes.
Defects make that the relative angles between the microcrystals that
form the polycrystalline structure are not arbitrary. For example,
panel (b) shows clearly that the relative angles between different
snub square tilings are multiple of $60$ degrees. The different vertex
that decorates the lattice of $M$-type particles are highlighted
with shaded plaquetes. Regions of the lattice that can not be joined
by regular polygons form defects. Representatives of them are highlighted
with red lines. Panel (c) shows that the structural motifs that decorate
the lattice of discs are globally different than the corresponding
to the $M$-type particles. However, the vertices that decorate the
lattice are of the same type. Also, a dodecagonal pattern usually
seen in quasicrystals is also highlighted. Red lines connects representative
defects on this lattice. Panel (d) summarizes the types of vertices
found in both lattices. The presence of the defects are relevant as
can be seen in the diffraction patterns {[}panels (e) and (f){]}.
A snub square tiling would produce a diffraction pattern with square
symmetry. However, the diffraction produced by the self-assembled
structure shows a pattern consistent with a twelve-fold symmetry.

\begin{figure}[H]
\centering{}\includegraphics[scale=0.5]{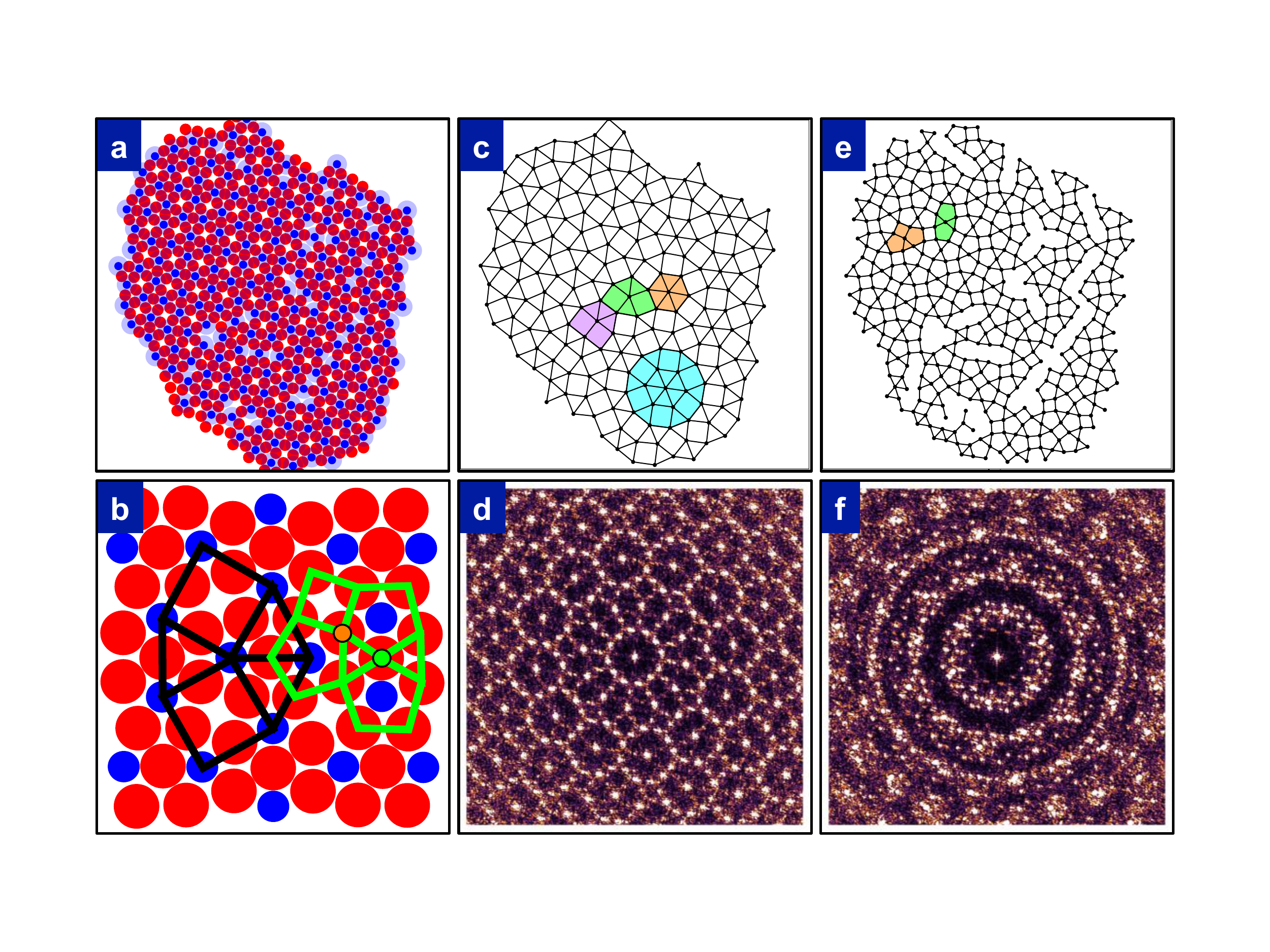}\caption{\textbf{\small Dodecagonal quasicrystal.}{\small{} Structure obtained
with $n=5$, $\sigma_{1}/\sigma_{0}$, as given by Eq. (1) and with
$\sigma_{2}/\sigma_{0}\simeq1.72$. (a) In this case a dodecagonal
quasicrystal of $M$-type particles (blue) is intercalated with a
lattice of discs forming pentagons (red).$ $ The lattice of $M$-type
particles shows a snub square crystal in the upper left quadrant.
The rest of the structure has a symmetry consistent with a twelvefold
symmetry. (b) Snub square section of the lattice, showing the $\left(3^{2}.4.3.4\right)$
mofit (black lines). The lattice of discs present two types of vertices,
a $\left(3.5.4.5\right)$ vertex (orange dot) and a $\left(3.5.3.5\right)$
vertex (green dot). These motifs are not formed by regular polygons
since the sum of their internal angles does not add to $360$ degrees.
Therefore, the polygons are slightly deformed (green lines). (c) Square-triangular
pattern corresponding to the $M$-type particles. Three neighbour
classification of $\sigma$ (green), $H$ (purple), and $Z$ (orange)
environments are shown. A dodecagonal motif typically found in quasicrystals
is highlighted in cyan. (d) Diffraction pattern of the lattice formed
by the $M$-type particles showing dodecagonal symmetry. (e) Pattern
of non-regular triangles and pentagons corresponding to the discs.
The two types of vertices are highlighted with orange $\left(3.5.4.5\right)$,
and green $\left(3.5.3.5\right)$. Regions where defects are present
can not be covered by these motifs. (f) Diffraction pattern of the
lattice of discs. The depths of the potential wells are $\left(1,1,1\right)$.}}
\label{suppfign5}
\end{figure}

A more detailed inspection of the case with $n=5$ is shown in Fig.
\ref{suppfign5}. Panel (a) shows a self-assembled lattice. Careful
examination of the structure shows a crystalline domain in the upper
left quadrant of the structure. This crystalline region is highlighted
in panel (b) where the structural motif of the lattice formed by $M$-type
particles is drawn with black lines. Again, a $\left(3^{2}.4.3.4\right)$
snub square tiling is formed. Lines connecting neighboring $M$-type
particles of the whole lattice show a square-triangular tiling {[}see
panel (c){]}. Formation of the dodecagonal quasicrystal in a square-triangle
lattice requires that the total tiling area occupied by squares be
equal to that occupied by triangles \cite{widom}, that is, $N_{3}/N_{4}=4/\sqrt{3}\simeq2.31$,
a value that closely corresponds with the simulation results. A confirmation
of this fact is the diffraction pattern of the $M$-type particles
lattice which is consistent with a dodecagonal quasicrystal, as shown
in panel (d). Panel (e) shows the polygonal tiling corresponding to
the positions of the discs. Two structural motifs are present, a $\left(3.5.3.5\right)$
vertex shown in green and a $\left(3.5.4.5\right)$ vertex shown in
orange {[}see also panel (b){]}. However, these motifs are not made
of regular polygons since the sum of their internal angles do not
add to $360$ degrees. Actually, a regular $n$-gon has internal angle
$\left(1-2/n\right)180$ degrees and there is a limited number of
combinations whose internal angles add to $360$ degrees. Thus, the
structural motifs of the lattice of discs are made of deformed polygons,
which are allowed thanks to the flexibility produced by the width
of the potential wells. The regions where the defects are present
can not be tessellated by these nearly regular polygons. The corresponding
diffraction pattern is shown in panel (f).

\newpage{}

\newpage{}

\end{document}